\documentclass{iaus}
\usepackage{graphicx}

\title[The core-degenerate scenario] 
{The core-degenerate scenario for type Ia supernovae}

\author[Noam Soker]   
{Noam Soker$^1$
}

\affiliation{$^1$Department of Physics, Technion -- Israel Institute of Technology, Haifa
32000, Israel \\ email: {\tt soker@physics.technion.ac.il}}

\pubyear{2008}
\volume{281}  
\pagerange{1--4}
\jname{Binary Paths to Type Ia Supernovae Explosions}
\editors{Rosanne Di Stefano \& Marina Orio, eds.}
\begin{document}

\maketitle

\begin{abstract}
In the core-degenerate (CD) scenario for the formation of Type Ia supernovae (SNe)
the Chandrasekhar or super-Chandrasekhar mass white dwarf (WD) is formed at the termination
of the common envelope phase or during the planetary nebula phase, from a merger of a WD companion
with the hot core of a massive asymptotic giant branch (AGB) star. The WD is destructed and
accreted onto the more massive core.
In the CD scenario the rapidly rotating WD is formed shortly after the stellar formation episode, and the delay from
stellar formation to explosion is basically determined by the spin-down time of the rapidly rotating merger remnant.
The spin-down is due to the magneto-dipole radiation torque.
Several properties of the CD scenario make it attractive compared with the double-degenerate (DD) scenario.
(1) Off-center ignition of carbon during the merger process is not likely to occur.
(2) No large envelope is formed. Hence avoiding too much mass loss that might bring the merger remnant
below the critical mass.
(3) This model explains the finding that more luminous SNe Ia occur preferentially in star forming galaxies.
\end{abstract}

\firstsection
\section{The Core Degenerate (CD) Scenario}

Observations and theoretical studies cannot teach us yet whether both the double-degenerate (DD) and
single degenerate (SD) scenarios for SNe Ia can work,
only one of them, or none (e.g., \cite{Livio2001}, \cite{Maoz2010}, \cite{Howell2011}).
I suggest to pay more attention to the \emph{core-degenerate} (CD) scenario that overcomes some difficulties in the
DD and SD scenarios (\cite{Ilkov2011, KashiSoker2011}, where more details can be found).

\begin{figure}[b]
\begin{center}
\includegraphics[scale=0.7]{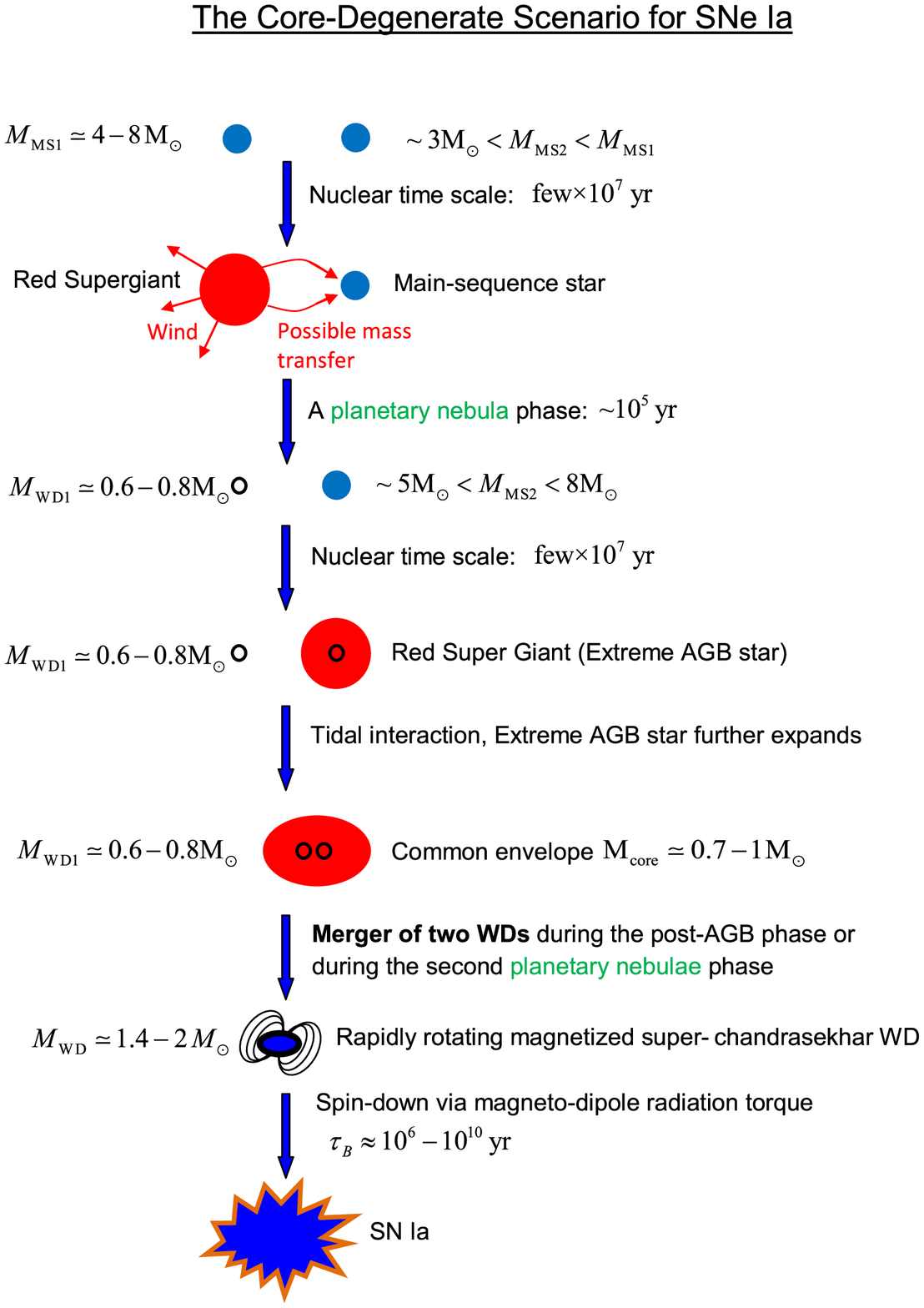}
        \caption{A schematic summary of the core-degenerate (CD) scenario for SNe Ia
        (from Ilkov \& Soker 2011).
        }
   \label{fig:fig1}
\end{center}
\end{figure}

The merger of a WD with the core of an AGB star was studied in the past
(\cite{Sparks1974}, \cite{Livio2003}, \cite{Tout2008}).
Livio \& Riess (2003) suggested that the merger of the WD with the AGB core
leads to a SN Ia that occurs at the end of the CE phase or shortly after, and can explain the presence of
hydrogen lines. In the CD scenario the possibility of a very long time delay (up to $10^{10}$~yr) is considered as well.
Because of its rapid rotation the super-Chandrasekhar WD does not explode (\cite{Yoon2005}).
The CD scenario is summarized schematically in Figure \ref{fig:fig1}.

Contrary to the view presented by Mario Livio in his review talk, I think the CD scenario is not a branch of
the DD scenario, but rather a distinguish scenario.
Both the CD and DD scenarios require the merger of the remnants of AGB stars (the core or the descendant WD)
to form a degenerate WD above the critical mass.
However, there are three key ingredients in the CD scenario that distinguish it from the DD scenario.
(1) The hot core is more massive than the companion cold WD.
(2) The merger should occur while the core is still large, hence hot. This limits the merger to occur within $\sim 10^5$~yr
after the common envelope phase. Kashi \& Soker (2011) showed that this condition can be met when the AGB star is massive.
(3) In the CD scenario most of the delay between the binary formation time and the explosion is due to the
spinning-down time of the merger product. The spinning-down is due to the magneto-dipole radiation
torque (and not gravitational waves; see \cite{Ilkov2011}).
In the DD scenario most of the delay time is the spiraling-in time of the two WDs (caused by gravitational radiation).

\firstsection
\section{The strong points of the CD scenario}

They most important factor is that the hot core is larger than its final radius when it becomes a cold WD.
At $\sim 10^5$~yr after it left the AGB the radius of a $M_{\rm core } \sim 0.7-0.8 M_\odot$ remnant is $\xi \simeq 1.2$
times its final radius as a cold WD (\cite{Bloecker1995}). This more or less limits the time period over which merger
must occur. Most likely the merger will occur much earlier, while the core is still large $\xi > 1.2$.
Since in the CD scenario the core is more massive than the WD companion, the WD companion will be destructed.

I now raise some strong points of the CD scenario, and compare it with the DD scenario.
\newline
\textbf{Carbon ignition off-center.}
The main problem for the DD scenario is that in many cases an off-center carbon ignition occurs
(e.g., \cite{SaioNomoto2004}) leading to accretion induced collapse (AIC) rather than a SNe Ia.
Yoon et al. (2007) raised the possibility that in a merger process where the more massive WD is hot,
off-axis ignition of carbon is less likely to occur.
The reason is that a hot WD is larger, such that its potential well is shallower
and the peak temperature of the destructed WD (the lighter WD) accreted material is lower.
Hence, in such a case the supercritical-mass remnant is more likely to ignite carbon in the center
at a later time, leading to a SN Ia.
Namely, the merger remnant becomes a rapidly rotating massive WD, that can collapse
only after it loses sufficient angular momentum.

\textbf{Mass loss of the merger product.}
Consider two merging cold WDs in the DD scenario. The less massive WD is destructed, and its mass is accreted
onto the more massive WD. The gravitational well of the more massive WD is much deeper than that of the
destructed WD (e.g., \cite{Dan2011}), and a large amount of energy is liberated $\sim 10^{50}$~erg.
If the remnant radiates the extra energy during a very short time $t_r$, we would expect for a very bright event
with a peak luminosity of $L_{\rm merg} \sim 10^8 (t_r/10~{\rm yr})^{-1} L_\odot$. This by itself
will be at an almost SN luminosity. Do we observe such objects?

If the energy release time is longer, the material of the destructed WD has time to expand and
form a giant-like structure (\cite{Shen2011}).
According to Shen et al. (2011) the giant-like phase lasts for $\sim 10^4$ years and its luminosity is
half the Eddington limit.
Such giants with a solar composition lose mass at a rate of few$\times 10^{-5} M_\odot~{\rm yr}^{-1}$ (\cite{Willson2007}).
When the carbon rich atmosphere of the merger remnant is considered the mass loss rate will be higher even.
Therefore, over the giant-like structure phase that lasts for $\sim 10^4 {\rm yr}$, the remnant might lose about half a solar mass
and decrease below the critical mass for explosion.

In the CD scenario the more massive WD is hot, and the potential well is much lower.
Assume a WD with a radius of $R_{\rm WD} \propto M_{\rm WD}^{-1/3}$ and a core with a radius of
$R_{\rm core} \propto \xi M_{\rm core}^{-1/3}$.
Then the ratio of the potentials is
\begin{equation}
\frac {\Psi_{\rm core}} {\Psi_{\rm WD}} \simeq
\frac{1}{\xi} \left( \frac {M_{\rm core}}{M_{\rm WD}} \right)^{4/3} =
1 \left( \frac{\xi}{1.5} \right)^{-1}
\left( \frac {M_{\rm core}/0.8M_\odot}{M_{\rm WD}/0.6M_\odot} \right)^{4/3} .
\label{eq:ed}
\end{equation}
The crude equality of potentials implies that the destruction of the less massive WD and
the accretion of its mass onto the core will not release large amount of energy,
and no formation of a giant-like structure will take place. The merger remnant will not have a large
radius, and no substantial mass loss will take place.
The merger remnant will continue to evolve as a massive central star of a planetary nebulae.

\textbf{More luminous SNe Ia in star forming galaxies}
The strong magnetic fields required in the present model for the spin-down mechanism most likely will enforce a rigid rotation within a short
time scale due the WD being a perfect conductor. The critical mass of rigidly rotating WDs is $1.48 M_\odot$
(\cite{Yoon2004} and references therein). This implies that WDs more massive than $1.48 M_\odot$
will explode in a relatively short time.
The similarity of most SN Ia suggests that their progenitors indeed come from a narrow mass range.
This is $\sim 1.4-1.48 M_\odot$ in the CD scenario.
This property of the magneto-dipole radiation torque spinning-down mechanism
explains the finding that SNe Ia in older populations are less luminous (e.g., \cite{Howell2001}; \cite{Smith2011}).


\end{document}